\documentclass[twocolumn,prb,showpacs,color,superscriptaddress,epsfig]{revtex4}
\usepackage{graphicx}
\usepackage{amsfonts}
\usepackage{amsmath}
\usepackage{amssymb}
\usepackage{epsfig}
\usepackage{color}
\begin{document}
\title{Exchange integrals and magnetization distribution in BaCu$_2$X$_2$O$_7$
(X=Ge,Si) }
\author{S. Bertaina}
\affiliation{Laboratoire Mat\'eriaux et Micro\'electronique de
Provence, Facult\'e St.\ J\'er\^ome, Case 142, F-13397 Marseille
Cedex 20, France} \affiliation{Laboratoire de Magn\'etisme Louis
N\'eel, CNRS, BP 166, 38042 Grenoble CEDEX-09,France}
\author{R. Hayn}
\affiliation{Laboratoire Mat\'eriaux et Micro\'electronique de
Provence, Facult\'e St.\ J\'er\^ome, Case 142, F-13397 Marseille
Cedex 20, France}

\date{\today}

\begin{abstract}
Estimating the intra- and inter-chain exchange constants in
BaCu$_2$X$_2$O$_7$ (X=Ge,Si) by means of density-functional
calculations within the local spin-density approximation (LSDA) we
find the Ge-compound to be a more ideal realization of a
one-dimensional spin chain with Dzyaloshinskii-Moriya interaction
than its Si-counterpart. Both compounds have a comparable magnitude
of inter-chain couplings in the range of $5 \ldots 10$ K, but the
nearest neighbor intra-chain exchange of the Ge-compound is nearly
twice as large as for the Si one. Using the LSDA+$U$ method we
predict the detailed magnetization density distribution and
especially remarkable magnetic moments at the oxygen sites.
\end{abstract}



\maketitle

The study of low-dimensional magnetism is of broad interest,
especially due to the crucial role of quantum fluctuations in those
systems. To investigate quantum spin chains experimentally one needs
low-dimensional model compounds. A representative example is
BaCu$_2$Ge$_2$O$_7$, \cite{bertaina04} or the isomorphous compound
BaCu$_2$Si$_2$O$_7$, \cite{tsukada99} knowing to be the most ideal
non-organic spin chain compound exhibiting the Dzyaloshinskii-Moriya
(DM)\cite{DM1,DM2} interaction.
 But why are some compounds more ideal than others?
What are the remaining differences to an ideal one-dimensional
compound? To answer these questions we present here a microscopic
calculation of the electronic structure of BaCu$_2$X$_2$O$_7$
(X=Ge,Si) within density-functional theory. We investigate the
chemical trends in comparing the Ge- with the Si-compound,
especially with respect to the intra-chain and inter-chain
tight-binding and exchange parameters. Furthermore, we calculate the
magnetization density in detail.

BaCu$_2$Ge$_2$O$_7$ and BaCu$_2$Si$_2$O$_7$ were first synthesized
and their crystallographic structures were described in 1993.
\cite{oliveira93} 
Both are composed of CuO$_4$ non-ideal plaquettes (the Cu atom and
the 4 O atoms are not perfectly coplanar) which are arranged into
corner-sharing chains along the $c$ axis (see Fig.\ 1) but with a
Cu-O-Cu bonding angle between 90 and 180 degrees. These bonding
angles are $\phi=135^{\circ}$ ($124^{\circ}$) for the
Ge(Si)-compound, and the zigzag chains allow for a DM interaction
(due to the low crystallographic symmetry, space group Pnma).
Correspondingly, both compounds were intensively studied as examples
for the influence of the DM interaction in quantum spin chains. A
weak ferromagnetism was observed in the Ge-compound \cite{tsukada00}
which allows the experimental determination of the spin canting
angle due to the DM interaction to be 1.9$^{\circ}$. On the other
hand, BaCu$_2$Si$_2$O$_7$ has attracted interest in connection with
the finding of two consecutive spin reorientation transitions.
\cite{tsukada01,zheludev02,glazkov} Also, two gaps in the excitation
spectrum of the Si-compound had been found by neutron scattering
\cite{kenzelmann01} and in antiferromagnetic resonance measurements.
\cite{hayn02} These two gaps arise due to the DM interaction and the
symmetric anisotropy term with a partial compensation between them.
\cite{hayn02,shekhtman92} Recently, it was the Ge-compound which
allowed to test a prediction of Oshikawa and Affleck
\cite{oshikawa02} in good accuracy. The linewidth of the electron
paramagnetic resonance signal diverges at low temperatures like
$\Delta H \propto 1/T^2$ for an applied magnetic field perpendicular
to the DM vector corresponding to the creation of a staggered field.
But for an applied field parallel to the DM vector, when no
staggered field can be created, the linewidth vanishes like $\Delta
H \propto T$. \cite{bertaina04}

\begin{figure}[htbp]
\includegraphics*[scale=0.45]{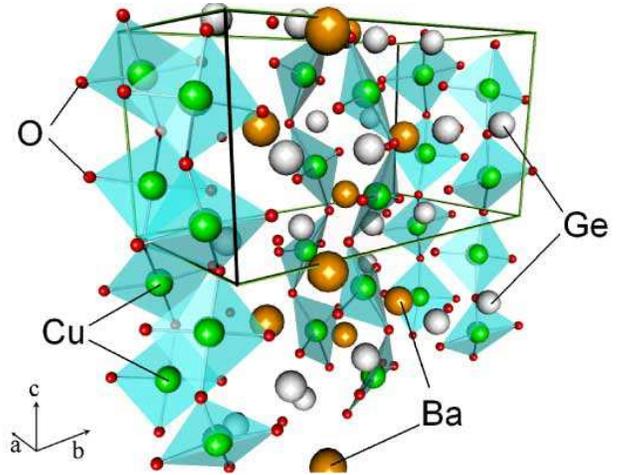}
\hspace{-12pt} \caption{(Color online) Crystal structure of
BaCu$_2$Ge$_2$O$_7$. The crystal structure is $Pnma$. The chains run
along the $c$ axis. } \label{fig:struc}
\end{figure}

Despite all these results, a microscopic calculation which
determines in detail the differences between the Si- and the
Ge-compound is lacking up to now. It will be presented here. Our
microscopic calculations confirm the naive expectation that the
difference in the bonding angles is the main reason that the
Ge-compound is a better 1D model system than the Si-one. In
addition, we also predict small oxygen moments in these compounds
which should be detectable by several experimental methods.

Our analysis of exchange couplings and model parameters follows the
lines given in Ref.\ \onlinecite{rosner97} for Sr$_2$CuO$_3$ and
Ca$_2$CuO$_3$.  We calculate the non-magnetic band-structure within
the local density approximation (LDA) and perform a tight-binding
(TB) analysis of the relevant band at the Fermi energy. Adding a
Hubbard interaction term we may propose an effective one-band
Hamiltonian which allows to estimate the different exchange terms.
These are in reasonable agreement with total energy differences from
the LSDA+$U$ method.

The scalar relativistic band-structure calculations were performed
using the full potential local orbital (FPLO) method.
\cite{koepernik99} In the FPLO method a minimum basis approach with
optimized local orbitals is employed, being at the same time a very
accurate and an efficient numerical tool. For the present
calculations we used the following basis set: Ba $4d5s5p$ : $6s6p$,
Cu $3s3p$ : $4s4p3d$, Ge $3s3p$ : $4s4p$ ; $3d5s$, Si $3s3p$, and O
$2s2p$ ; $3d$. The inclusion of the Ba $4d5s5p$ and the Cu $3s3p$
semi-core states into the valence set was necessary to account for
non-negligible core-core overlap, and the O $3d$ states were used to
improve the completeness of the basis set. The site-centered
potentials and densities were expanded in spherical harmonics up to
$l_{max}=12$. The LDA calculations were performed using the
Perdew-Wang parametrization. \cite{perdew92} In all calculations we
used a $k$-mesh of $4\times4\times4=64$ $k$-points in the Brillouin
zone which is sufficient for the band-structure and magnetization
density calculations reported below.

\begin{figure}[htbp]
\includegraphics*[bb=38 70 529 763,angle=-90,scale=0.35,clip]{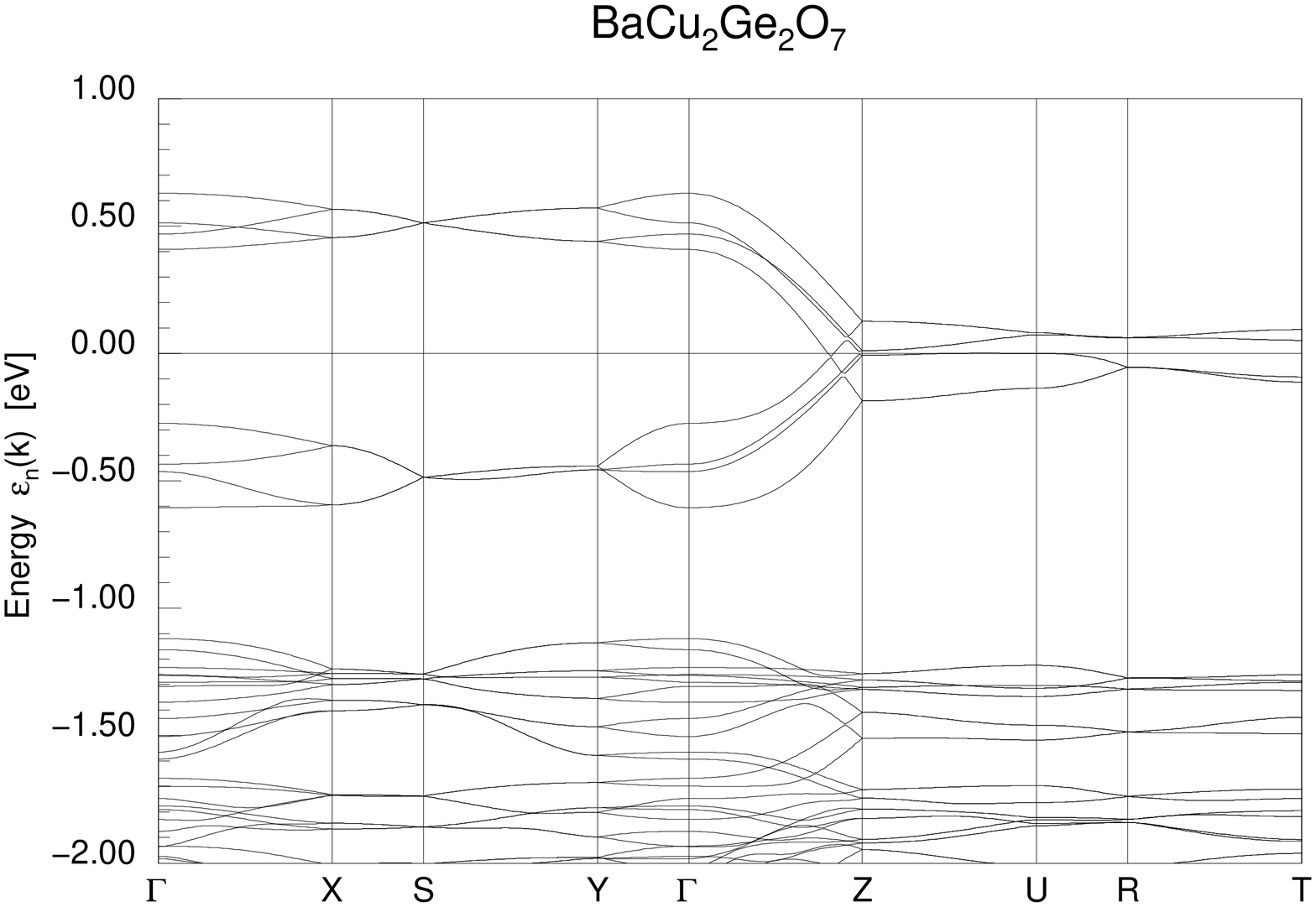}
\hspace{-12pt} \caption{Band-structure of BaCu$_2$Ge$_2$O$_7$ within
a non-magnetic state using LDA. The momentum points in the Brillouin
space are given in units of ($\pi/a,\pi/b,\pi/c$) by
$\Gamma=(0,0,0)$, $X=(1,0,0)$, $Y=(0,1,0)$, $Z=(0,0,1)$,
$S=(1,1,0)$, $U=(1,0,1)$, $R=(1,1,1)$ and $T=(0,1,1)$. $\Gamma
\rightarrow Z $ is the dispersion along the chain axis. }
\label{fig:band}
\end{figure}

The non-magnetic band-structure of BaCu$_2$Ge$_2$O$_7$ is shown in
Fig.\ 2. One may note a well isolated band at the Fermi level having
8 branches according to the 8 Cu ions per elementary cell (Fig.\ 1).
The main valence band starts at about -1 eV and extends up to -7 eV.
It is comprised of highly hybridized Cu $3d$ and O $2p$ orbitals as
it is common for cuprate compounds. \cite{rosner97} The band at the
Fermi level consists mainly of the Cu $3d_{x^2-y^2}$ orbital lying
predominantly within the plane of the CuO$_4$ plaquette (in the
local plaquette coordinate system) and hybridizing slightly by
$pd\sigma$ bonds with the oxygen $2p$ orbitals directed towards Cu.
One has to choose the non-magnetic solution to determine the TB
parameters \cite{rosner97} since the Hubbard correlation will be
added later on by hand. Taking into account nearest neighbor hopping
processes into all 3 cartesian directions and providing a
down-folding into the present Brillouin zone we may approximately
describe the 8 branches of the dispersion of the relevant band by
\begin{equation}
\label{eq1} E_k=\pm 2 t_z \mbox{cos} \left( \frac{k_z c}{2} \right)
\pm 2 t_x \mbox{cos} \left( \frac{k_x a}{2} \right) \pm 2 t_y
\mbox{cos} \left( \frac{k_y b}{2} \right) \; .
\end{equation}
The corresponding fit gives the parameters listed in the Table
\ref{tab:comp}. The errors are estimated by the evident deviations
of the real band-structure (Fig.\ 2) from Eqn.\ (\ref{eq1}). They
correspond to additional hopping processes not taken into account in
the simple model with only nearest-neighbor hopping. Comparing the
Si- with the Ge-compound we may note a rough agreement with respect
to the inter-chain hoppings. But, as can be also expected from the
different bonding angles $\phi$, the in-chain hopping is clearly
enhanced for BaCu$_2$Ge$_2$O$_7$ in comparison to its
Si-counterpart.

\begin{table}[htbp]
\begin{tabular}{lcc}
\hline
  & BaCu$_2$Ge$_2$O$_7$  & BaCu$_2$Si$_2$O$_7$  \\
\hline \hline $\phi$ & 135$^{\circ}$ & 124$^{\circ}$ \\
$4t_z=4t$& (950$\pm$20)meV & (700$\pm$45)meV \\
$4t_y$& (130$\pm$15)meV & (85$\pm$20)meV \\
$4t_x$& (130$\pm$30)meV & (150$\pm$30)meV \\
\hline
$J_{exp}=J$&540K=46.5meV\cite{tsukada00}& 280K=24.1meV\cite{tsukada99} \\
$T_{N,exp}=T_N$& 8.6K\cite{tsukada00}& 9.2K\cite{tsukada99} \\
$J_{\perp}^{emp}$& 3.6K & 4.1K \\
\hline
$U$ & 4.8eV & 5.1eV\\
$t_{\perp}$ & 110 \ldots 150meV & 100 \ldots 140meV\\
$J_{\perp}^{band}$&7.3 \ldots 13.6K&5.7 \ldots 11.1K\\
\hline
  \end{tabular}
  \caption{Collection of experimental data and theoretical model
  constants for BaCu$_2$X$_2$O$_7$ (X=Ge,Si). }
  \label{tab:comp}
\end{table}

We may use the derived TB parameters to construct a highly
correlated one-band Hubbard model
\begin{eqnarray}
\label{eq2} H=&& t\sum_{i,\pm z, \sigma} c_{i, \sigma}^+c_{i\pm z,
\sigma} + t_{\perp} \sum_{i,\pm x/y, \sigma} c_{i, \sigma}^+c_{i\pm
x/y, \sigma} \nonumber \\ &&+ U \sum_i n_{i,\downarrow}
n_{i,\uparrow}
\end{eqnarray}
by identifying $t_z=t$ and using for $t_{\perp}$ the average value
of $t_x$ and $t_y$. This Hubbard-like Hamiltonian may be further
reduced to a Heisenberg one $H=(1/2) \sum_{i,j} J_{ij} \bf{S}_i
\bf{S}_j$, where the exchange couplings are connected with the TB
parameters by $J_{ij}=4t_{ij}^2/U$. Please note, that for the
purpose of estimating the inter-chain couplings, we neglected those
terms in the Hamiltonian which are the most interesting in the
present case, namely the DM term and the symmetric anisotropy. They
are caused by the spin-orbit coupling $\lambda \bf{S} \bf{L}$
($\lambda \approx 0.1$ eV) and should be understood to be added to
the isotropic part. A perturbative estimate leads to an orbital
moment $ \propto \lambda/\varepsilon_d$ ($\varepsilon_d$ - crystal
field splitting of $d$-levels) and a spin canting angle of
$3^{\circ}$ caused by the DM interaction $\propto \lambda$. The
symmetric anisotropy $\propto \lambda^2$ compensates partly the DM
term and it is just the Hund's rule coupling which leads to a
"residual anisotropy" and to the two gaps in the magnetic excitation
spectrum. \cite{hayn02}

Experimental information about the nearest-neighbor exchange
coupling $J$ was derived from the maximum of the susceptibility
curves \cite{tsukada99,tsukada00} using the Bonner-Fisher theory
\cite{bonner64} (see Table \ref{tab:comp}). Using simply the same
$U$ for both compounds one would predict the ratio
$J^{Ge}/J^{Si}=1.84$ from the theoretical $t=t_z$ values, which is
already in good agreement with the experimental ratio 1.9. One can
also take the experimental $J$-values and the band-structure $t$'s
to derive an effective one-band Hubbard $U$ of 4.8 eV (Ge) and 5.1
eV (Si) which seem to be quite reasonable (see Table
\ref{tab:comp}). In the given case the simple formula $J=4 t^2/U$
works much better than for Sr$_2$CuO$_3$ or CuGeO$_3$ where
considerable ferromagnetic corrections are necessary.
\cite{rosner97} Due to the zigzag chains the hopping $t$ parameters
of Table \ref{tab:comp} are much smaller than for Sr$_2$CuO$_3$
which gives a better justification of the second order perturbation
theory. On the other hand, the Cu ions are more far away from each
other than in CuGeO$_3$ which reduces the direct ferromagnetic
exchange contributions.

The theoretical inter-chain hopping and exchange parameters are
connected with quite important error bars (which is, however, not
unusual for cuprate compounds). One possibility to derive the
magnetic inter-chain couplings was discussed above and leads to
$J_{\perp}^{band} = 4 t_{\perp}^2 / U$ (see Table \ref{tab:comp}).
Another method uses the theory of Irkhin and Katanin \cite{irkhin00}
which is an improved Schulz theory \cite{schulz96} of weakly coupled
spin chains. It was recently confirmed by detailed quantum Monte
Carlo calculations. \cite{Yasuda05} In the theory of Irkhin and
Katanin, the N\'eel temperature $T_N$ is given by
\begin{equation}
\label{eq3} T_N= k J_{\perp} z_{\perp} \tilde \chi_0 L(\Lambda J /
T_N)
\end{equation}
with the numerical constants $k=0.70$, $\tilde \chi_0=2.1884$,
$\Lambda=5.8$, the number of nearest neighbors $z_{\perp}=4$, and
the scaling function
\begin{equation}
\label{eq4} L(\Lambda J / T_N)= B \left[ \mbox{ln} \frac{\Lambda
J}{T_N} + \frac{1}{2} \mbox{ln} \;  \mbox{ln} \frac{\Lambda J}{T_N}
\right]^{\frac{1}{2}}
\end{equation}
with $B=0.15$. Applying (\ref{eq3}) and (\ref{eq4}) and using the
measured values of $T_{N,exp}$ and $J_{exp}$ we derived the
empirical inter-chain exchange constants $J_{\perp}^{emp}$ collected
in Table \ref{tab:comp}. Both methods lead to more or less identical
inter-chain couplings for the two compounds. But the
$J_{\perp}^{band}$ are roughly a factor of two larger than the
empirical values $J_{\perp}^{emp}$. A similar discrepancy had been
already observed for Sr$_2$CuO$_3$, Ca$_2$CuO$_3$ and CuGeO$_3$ in
Ref.\ \onlinecite{rosner97}, where also possible origins of that
discrepancy had been discussed. Our present estimate gives only a
rough idea about the absolute values of inter-chain couplings, but
cannot answer whether these couplings would be ferro- or
antiferromagnetic. Certainly, a more precise determination of
inter-chain couplings deserves more investigations.

\begin{figure}[htbp]
\includegraphics*[bb=15 15 285 221,scale=0.9,clip]{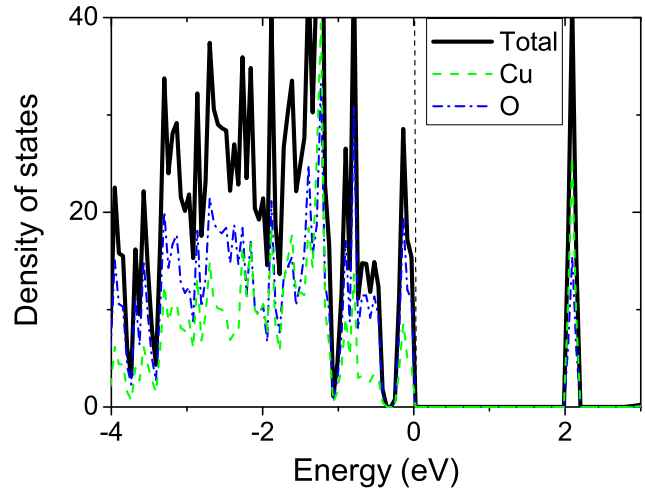}
\hspace{-12pt} \caption{(Color online) Total and partial density of
states (DOS) of BaCu$_2$Ge$_2$O$_7$ from LSDA+$U$ calculation. The
thick black line is the total DOS whereas the thin green and blue
lines are the DOS of all coppers and oxygens orbitals,
respectively.} \label{fig:dos}
\end{figure}

\begin{table}
\begin{center}
\begin{tabular}{p{2cm} p{1cm} p{1cm} p{1cm} p{1cm}}
\hline\hline
  & $m_{Cu}$ & $m_{O1}$ & $m_{O2}$ & $m_{O3}$ \\
\hline LSDA & & & & \\
BaCu$_2$Ge$_2$O$_7$  & 0.291 & 0.062 & 0.070 & 0.002 \\
BaCu$_2$Si$_2$O$_7$ & 0.478 & 0.067 & 0.066 & 0.002 \\
\hline LSDA+$U$ & & & & \\
BaCu$_2$Ge$_2$O$_7$  & 0.623 & 0.054 & 0.082 & 0.009 \\
BaCu$_2$Si$_2$O$_7$ & 0.680 & 0.062 & 0.061 & 0.002 \\
\hline \hline \hline
\end{tabular}
\caption{Distribution of magnetization in BaCu$_2$X$_2$O$_7$ with
X=Ge,Si using FPLO within the LSDA and the LSDA+$U$ schemes.
Numerical values are in Bohr magneton.} \label{tab:mag}
\end{center}
\end{table}

By allowing an antiferromagnetic (AFM) arrangements of spins we
found an insulating AFM solution already within the local
spin-density approximation (LSDA). However, the obtained gap values
of about 0.5 eV are unrealistically small in LSDA. Therefore, we
performed LSDA+$U$ calculations with the Slater-Coulomb parameters
$F_2$=8.6eV, $F_4$=5.4eV (corresponding to $J=1$ eV) and two values
for $F_0=U=3.7$ eV and 5 eV. The former value produces gaps
corresponding to the deep blue color of the samples (Fig.\
\ref{fig:dos}), but the latter one is more close to parameters
common for cuprates. \cite{fn1} We used the same optimized basis
orbitals like for the LSDA calculation and the "around mean field"
version of LSDA+$U$. The total energy differences between FM and AFM
solution per magnetic ion $\Delta E$ lead to the nearest neighbor
exchanges $J=\Delta E / \mbox{ln} 2$ of 46.6 (65.2) meV and 22.6
(45.5) meV for the Ge- and Si-compound with $U=$ 5.0 (3.7) eV.


\begin{figure}
\includegraphics*[bb=19 18 525 416,width=0.45\textwidth,clip]{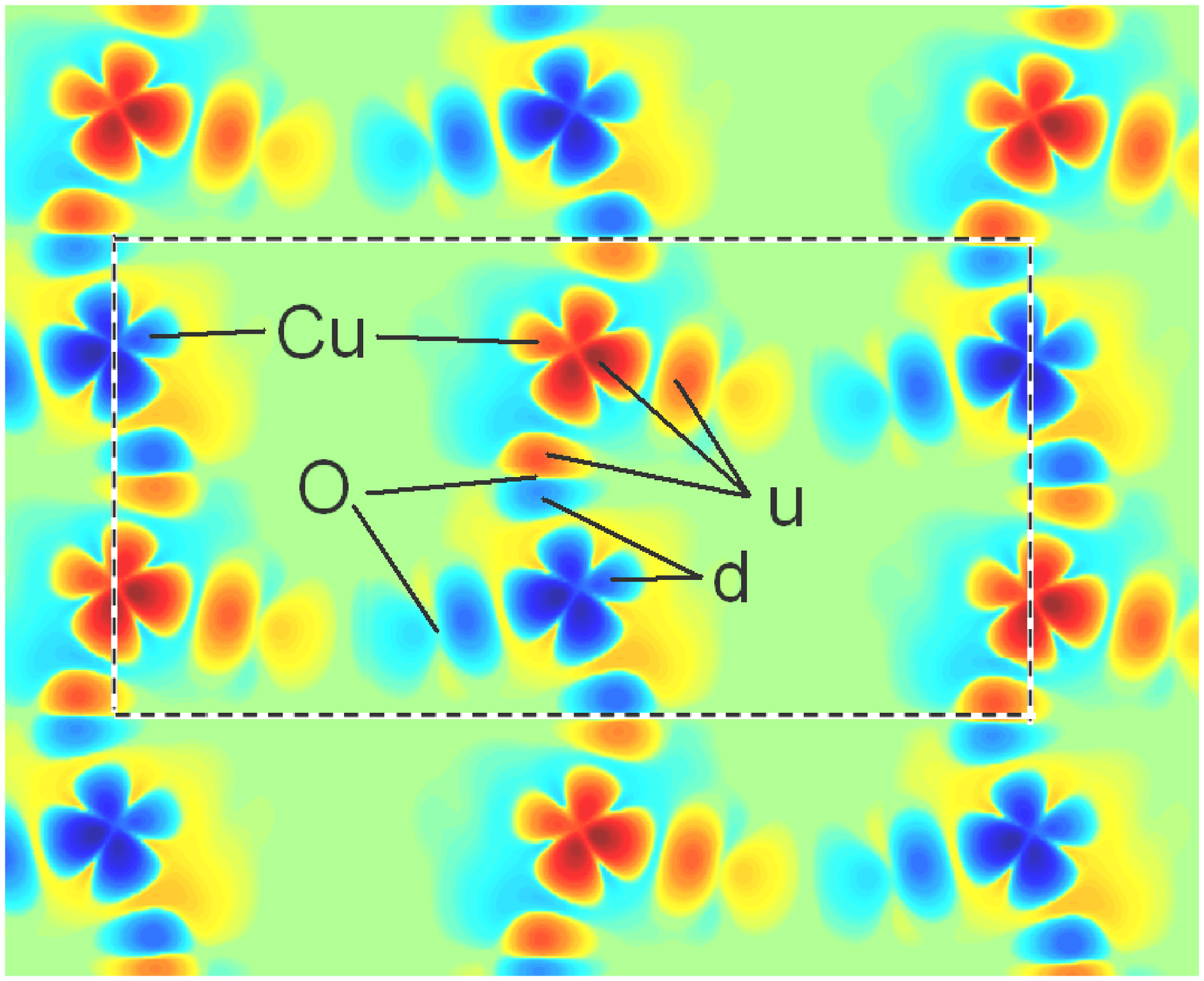}
\caption{(Color online)Local magnetization distribution of
BaCu$_2$Ge$_2$O$_7$ in the $bc$ plane passing through the vertical
chain axis $c$. Shown is the spin density in a logarithmic scale:
yellow to red (blue) parts correspond to spin up [u] (down [d])
density, green parts are very low local density. Dashed lines
represent the elementary cell.} \label{fig:dens}
\end{figure}

The magnetization distribution on the different sites of the two
compounds within LSDA and LSDA+$U$ (for $F_0=3.7$ eV having only
small deviations to the $F_0=5$ eV solution) is shown in Table
\ref{tab:mag}. \cite{remarque2} In the plaquette, each Cu ion is
surrounded by 4 oxygens: O1 and O2 are the side oxygens, and O3 the
bridging oxygen between two plaquettes. As one can see, a non
negligible part of the magnetization is situated at the oxygen
sites, mainly at O1 and O2.
Compared to the LSDA results, the proper inclusion of Coulomb
correlations leads to an increase of magnetization at the Cu sites
since it stabilizes the AFM long range order by decreasing the
quantum fluctuations.

To see the magnetization distribution in more detail we calculated
the spin density $n_\uparrow-n_\downarrow$ in real space within the
elementary cell. It is presented in Fig.\ \ref{fig:dens} in a
logarithmic scale. We neglect here the small deviations from a
collinear spin arrangement, which is justified by the small spin
canting angle of $1^{\circ} \ldots 2^{\circ}$. One observes the
characteristic form of $3d$ orbitals at the Cu sites and also the
different oxygen sites. Quite remarkable are the different
magnetization directions in the two lobes of the bridging oxygen,
which explains the small numbers for $m_{O3}$. A sign change of the
magnetization at O3 is theoretically possible by $sp$ hybridization.
On the contrary, the magnetization points into one direction at the
side oxygens.


Summarizing, we determined model parameters and predicted oxygen
moments in BaCu$_2$X$_2$O$_7$ (X=Ge,Si). The absolute values of
inter-chain couplings $|J_{\perp}|$ are very similar for
BaCu$_2$Ge$_2$O$_7$ and BaCu$_2$Si$_2$O$_7$, but the nearest
neighbor intra-chain $J$'s are different. The larger $J$ for the
Ge-system makes it to be a more ideal model compound than
BaCu$_2$Si$_2$O$_7$.
Our calculated oxygen moments should be observable by $^{17}$O
nuclear magnetic resonance, by $\mu$-spin rotation measurements, or
by neutron scattering. They are similar (but slightly smaller) than
those in Li$_2$CuO$_2$ which were recently detected. \cite{Zheludev}


We thank A. Stepanov and H. Rosner for interesting discussions.


\begin{thebibliography}{30}
\bibitem{fn1}
Please note that for the "standard" parameter set with $U=5$ eV the
lowest unoccupied band of our calculation is of Cu $4s$ character in
contrast to the expected Cu $3d$ orbital.
\end{thebibliography}
\end{document}